\input amstex
\documentstyle{amsppt}
\magnification 1200
\NoBlackBoxes
\vcorrection{6pt}
\hcorrection{20pt}
\pageheight{8.5 true in}
\pagewidth{6 true in}
\topmatter
\parskip 4pt
\title
On decomposing $N=2$ line bundles \\ as tensor products of $N=1$ line
bundles
\endtitle
\rightheadtext{Tensor decomposition of $N=2$ line bundles}
\author Fausto Ongay$^1$ and Jeffrey M. Rabin$^2$ \endauthor 
\leftheadtext{F. Ongay and J. M. Rabin}
\affil 
$^1$CIMAT, Guanajuato, Gto. 36240 Mexico\\ 
$^2$Dept. of Mathematics, UCSD, La Jolla CA 92093 USA\\e-mail:
 ongay\@cimat.mx; jrabin\@ucsd.edu
\endaffil 
\abstract 
We  obtain the existence of a
cohomological obstruction to expressing $N=2$ line bundles as tensor
products of $N=1$ bundles. The motivation behind this paper is an attempt
at understanding the
$N=2$ super KP equation via Baker functions, which are special sections
of line bundles on supercurves.
\endabstract
\thanks 
Partially supported by CONACYT, Mexico, project 37558-E. 
\endthanks
\endtopmatter
\document


There has been---for some time now (cf. \cite{DG})---an interest in
extending the study of the super KP equations from the case $N=1$ to the
case $N=2$. 
One possible way to do
this, that could be particularly useful for understanding the geometry
of these equations, would be via Baker functions: Roughly
speaking, Baker functions are  special unique sections with parameters of
certain families of line bundles, satisfying the condition that any
section of the corresponding line bundle is given by a differential
operator applied to the Baker function. From the geometric point of view,
their relevance stems from the fact that they allow us to reinterpret
equations such as the KP (or its super analogs), as describing
deformations of line bundles over curves (or supercurves). 

Thus, as a necessary step towards such a study of the super KP equations,
one must first understand the geometry of $N=2$ superline bundles, and this
note is aimed towards that goal. Specifically, what we obtain here is the
existence of a cohomological obstruction to expressing $N=2$ line bundles
as tensor products of $N=1$ bundles. (And therefore,  in
general there might not be a simple relationship between
$N=1$ and $N=2$ Baker functions.)

\vskip 6pt
To properly describe our result, let us first recall that a
smooth (complex) supercurve
$M$ is a family of ringed spaces
$(M_{\text{red}},\Cal{O}_M)$, over the base $(\bullet,\Lambda)$. Here
$\Lambda$ is a Grassmann algebra (in some generators, say
$\eta_1,\dots,\eta_n$),
$M_{\text{red}}$ is an ordinary smooth complex curve, and
the structure sheaf $\Cal{O}_M$ is a sheaf of
$\Bbb{Z}_2$-graded algebras, locally isomorphic (as sheaves of
$\Bbb{Z}_2$-graded algebras) to
$\Cal{O}_{\text{red}}\otimes\Lambda[\theta^1,\dots,\theta^N]$,
$\Cal{O}_{\text{red}}$ being the structure sheaf of  $M_{\text{red}}$, and
$\theta^\alpha$ odd generators nilpotent of order two. (Super)line bundles
over the supercurve are then defined as locally free sheaves of rank 1
$\Cal{O}_M$-modules. 

Supercurves can be described in a more concrete way, by prescribing
the changes of coordinates between charts: for the case
$N=1$, where we simply write $\Cal{O}=\Cal{O}_M$, 
this takes on the form
$$\align
z_j=&\ f_{ji}(z_i)+\theta_i\gamma_{ji}(z_i)\\
\theta_j=&\ \mu_{ji}(z_i)+\theta_in_{ji}(z_i),
\endalign$$
where $f_{ji},\ n_{ji}$ are even invertible holomorphic functions, while
$\gamma_{ji},\
\mu_{ji}$ are odd.
Line bundles over $M$ (i.e., $N=1$ bundles) are then determined by
transition functions
$\Gamma_{ji}=a_{ji}+\theta_i\alpha_{ji}$, with $\Gamma_{ji}$ an invertible
superfunction satisfying a cocycle condition.

To any such supercurve $M$ there is an associated $N=2$ super Riemann
surface (SRS), denoted
$M_2$, constructed by adjoining an odd coordinate $\rho$ that
transforms according to
$$\rho_j=\frac{\gamma_{ji}}{n_{ji}}-
\theta_i\rho_i\left(\frac{\gamma_{ji}}{n_{ji}}\right)'+
\rho_i\frac{f_{ji}'n_{ji}-\mu_{ji}'\gamma_{ji}}{n_{ji}^2},$$
the coefficient of $\rho$ being the Berezinian of the change of
coordinates on $M$; we denote the structure sheaf of $M_2$ by
$\Cal{O}_2$. Moreover, by considering $\rho_i$ as the odd
coordinate and $\hat z_i=z_i-\theta_i\rho_i$ as the
even coordinate one gets, associated to $M$, another $N=1$ supercurve,
$\hat M$, called the {\it dual\/} supercurve, whose structure sheaf we
denote as
$\hat{\Cal{O}}$. 

Recall that any $N=2$ super Riemann
surface has a global rank
$0\vert2$ nonintegrable distribution, locally generated by
$D^+=\partial_\rho$ and
$D^-=\partial_\theta+\rho\partial_z$.
However, $M_2$ as constructed above is always an ``untwisted" $N=2$
SRS, meaning that each of $D^\pm$ generates a rank $0\vert1$
distribution.  The sections annihilated by
$D^+$ are called {\it chiral\/} sections, those annihilated by $D^-$
{\it antichiral\/}.
Moreover, the structure sheaf of $M_2$ fits into the exact sequence
$$0\to\Cal{O}\to\Cal{O}_2\to\Cal{B}er\to0,\tag{1}$$
where $\Cal{B}er$ denotes the Berezinian sheaf of $M$, and there is a
similar sequence for $\hat{M}$, namely
$$0\to\hat\Cal{O}\to\Cal{O}_2\to\hat\Cal{B}er\to0.$$
The corresponding projections in these sequences are just the operators
$D^+$ and
$D^-$, respectively.


Observe that a line bundle over either $M$ or $\hat{M}$ can be viewed
as a line bundle over $M_2$ having the same transition functions (these
are the pullbacks via the obvious projections $M_2\to M$ and $M_2\to
\hat{M}$), and their tensor product is a line bundle over $M_2$ as
well. Conversely, given a line bundle on $M_2$ we can ask when it splits as
a tensor product of line bundles on
$M$ and $\hat M$.

To answer this question, first of all we make the observation that any
$N=2$ local superfunction, 
$$\Gamma=h+\theta\phi+\rho\psi+\theta\rho g,$$
can be decomposed as a product of a function on $M$ and a function
on $\hat M$, i.e., 
$$\Gamma=\big(a(z)+\theta\alpha(z)\big)\big(b(\hat
z)+\rho\beta(\hat z)\big).$$ 
In fact,
$$h=ab\;;\ \phi=\alpha b\;;\ \psi=a\beta\;;\ g=-(ab'+\alpha\beta),$$
where now all functions depend on $z$. Since we can obtain
$b'/b$ from them, these conditions actually determine $b$ up to a
multiplicative constant (and $\beta$ also); $a$ is then
determined up to the {\it reciprocal\/} constant (and so is $\alpha$).

\vskip 6pt
{\bf Remark}. There is a similar decomposition of
$N=2$ local superfunctions as sums of $N=1$ superfunctions,
$\Gamma(z,\theta,\rho)=F(z,\theta)+\hat{F}(\hat{z},\rho)$,
modulo an additive constant. This, of course, is essentially ``taking the
logarithm" of the decomposition above, and is a manifestation of the
existence of an exponential exact sequence
$$0\to\Bbb{Z}\to\Cal{O}\to\Cal{O}^\times\to0,$$
where, as usual, $^\times$ denotes the invertible elements; however, we
will not need this construction in what follows.

\vskip 6pt
Now, if $\Gamma_{ji}$ are the
transition functions of an $N=2$ line bundle, written as above, since the
$a_{ji}$ and $b_{ji}$ themselves are only determined up to a constant, say
$c_{ji}$, what this actually gives is the existence of a short exact
sequence of the form
$$
0\to\Lambda_{\text{ev}}^{\times}\to
\Cal{O}_{\text{ev}}^{\times}\times\hat{\Cal{O}}^{\times}_{\text{ev}}\to
\Cal{O}^{\times}_{2,\text{ev}}\to0,\tag{2}$$
 where
$\Cal{O}_{\text{ev}}^{\times}\times\hat{\Cal{O}}_{\text{ev}}^{\times}$ is
the sheaf described at the presheaf level by taking direct products of the
groups
$\Cal{O}_{\text{ev}}^\times(U_i)$ and
$\hat\Cal{O}_{\text{ev}}^\times(U_i)$. The second arrow is the map
$c\mapsto(c,c^{-1})$ and the third is
$(F,\hat{F})\mapsto F\hat{F}$. Notice that all objects appearing in the
construction of the sequence above are {\it even\/}; in particular the
sheaves are sheaves of abelian groups.

We can now state our main result:

\proclaim{Theorem} Let $M$ be an $N=1$ supercurve. Then for any
given line bundle $\Cal{L}$ over the associated $N=2$ SRS $M_2$, there
exists a cohomology class in
$H^2(M_{\text{red}},\Lambda_{\text{ev}}^\times)$, that measures the
obstruction to expressing  $\Cal{L}$ as a tensor product of line bundles
over
$M$ and $\hat{M}$.
\endproclaim

\noindent{\it Proof\/}: Recall that, in general, line bundles over a
complex  (super) manifold are
classified by the first cohomology group,
$H^1(M_{\text{red}},\Cal{O}_{\text{ev}}^{\times})$. 

Now, taking
the cohomology sequence of the short exact sequence (2) above, the last
part reads
$$\dots\to
H^1(M_{\text{red}},\Cal{O}_{\text{ev}}^{\times}\times
\hat{\Cal{O}}_{\text{ev}}^{\times})
\to
 H^1(M_{\text{red}},\Cal{O}^{\times}_{2,\text{ev}})@>\beta>>
H^2(M_{\text{red}},\Lambda^{\times})\to0 .\tag{3}$$
However, there is a canonical map of
cohomology groups
$$H^1(M_{\text{red}},\Cal{O}_{\text{ev}}^{\times})\times
H^1(M_{\text{red}},\hat{\Cal{O}}_{\text{ev}}^{\times})\to
H^1(M_{\text{red}},\Cal{O}_{\text{ev}}^{\times}\times
\hat{\Cal{O}}_{\text{ev}}^{\times}),$$ 
and this map is in fact an isomorphism, because the domain and range are
both defined by  equivalence relations on pairs of
cocycles of transition functions. Indeed, the isomorphism is (essentially)
determined by the assignment
$\big(F(z),\hat{F}(\hat{z})\big)\mapsto\big(F(z),\hat{F}(z)-\theta\rho
\hat{F}'(z)\big)$.

Therefore, we have a
group morphism 
$$\delta\:
H^1(M_{\text{red}},\Cal{O}_{\text{ev}}^{\times})\times
H^1(M_{\text{red}},\hat{\Cal{O}}_{\text{ev}}^{\times})\to
H^1(M_{\text{red}},\Cal{O}_{2,\text{ev}}^{\times}),
$$
whose image, by construction, consists of the isomorphism classes of
$N=2$ bundles that can be expressed as tensor product of bundles over
$M$ and $\hat{M}$. Thus, a line bundle over $M_2$ can be decomposed as a
tensor product  if and only if the corresponding class is in the image of
$\delta$, which equals the kernel of $\beta$ in (3); this gives
the cohomology obstruction whose existence was asserted.
\qed

\vskip 6pt
Let us make some remarks in relation to this result. 

First of all, a somewhat more
concrete proof of the theorem can be given along the following lines:
Given a cocycle of transition functions for an
$N=2$ line bundle
$\Cal{L}$, recalling the sequence (1), we can form the quotients
$$\frac{D^+\Gamma_{ji}}{\Gamma_{ji}}\ ;\quad
\frac{D^-\Gamma_{ji}}{\Gamma_{ji}},$$
and these are in fact sections of the Berezinian sheaves $\Cal{B}er(M)$ and
$\hat\Cal{B}er(M)$ of $M$ and $\hat{M}$ respectively. By indefinite integration of these local sections and then
exponentiation,  up to multiplicative constants one gets functions
$F_{ji}$ and
$\hat{F}_{ji}$, and this gives a factorization of the form
$\Gamma=F\hat{F}$ for the transition functions of $\Cal{L}$  (see
\cite{BR} for details). Now, the
cocycle conditions for
$\Gamma_{ji}$ show that on triple overlaps one has 
$$\log{F_{ji}}+\log{F_{kj}}+\log{F_{ki}}=
\log{\hat{F}_{ji}}+\log{\hat{F}_{kj}}+\log{\hat{F}_{ki}},$$ 
modulo some integers (representing the Chern class of the bundle
$\Cal{L}$). But the left hand side is chiral while the right hand side is
antichiral, so that they are necessarily equal to a constant, say
$c_{kji}$; then
$\exp{c_{kji}}$ gives an explicit representative for the desired class
in
$H^2(M_{\text{red}},\Lambda_{\text{ev}}^\times)$.

On the other hand, the proof  given above 
sheds some light on an interesting phenomenon. Namely, it was known that
for certain supercurves (nonprojected generic SKP curves, cf. below) there
are examples of nontrivial
$N=1$ bundles
$\Cal{L}$ giving a nontrivial factorization
$\Cal{O}_2=\Cal{L}\otimes\hat\Cal{O}$, in addition to the trivial one,
$\Cal{O}_2=\Cal{O}\otimes\hat\Cal{O}$; that is, $\Cal{L}$ is
nontrivial as a bundle over $M$, but its lift to $M_2$ is trivial
\cite{BR}. 

This can be explained as follows: 

By going one further step backwards
in the cohomology sequence (3), we get an exact sequence of the form 
$$\dots\to H^1(M_{\text{red}},\Lambda_{\text{ev}}^\times)\to
H^1(M_{\text{red}},\Cal{O}_{\text{ev}}^{\times})\times
H^1(M_{\text{red}},\hat{\Cal{O}}_{\text{ev}}^{\times})\to
H^1(M_{\text{red}},\Cal{O}_{2,\text{ev}}^{\times})\to\dots$$ 
By construction, the second arrow in this sequence maps a flat line
bundle (or, to be precise, its isomorphism class), say
$\Cal{L}$, defined by the cocycle of constant transition functions
$c_{ji}$, to the pair of bundles $(\Cal{L},\Cal{L}^{-1})$, defined by the
pair
$(c_{ji},c_{ji}^{-1})$, where the bundles of the pair are regarded as
being over
$M$ and $\hat{M}$ respectively. 

Thus, the assertion above is that the
cocycle
$c_{ji}$ might be trivial when seen as representing an element of 
$H^1(M_{\text{red}},\Cal{O}_{\text{ev}}^{\times})$ (i.e., as
defining a bundle over $M$), {\it but not\/} when seen as
an element of
$H^1(M_{\text{red}},\hat{\Cal{O}}_{\text{ev}}^{\times})$ (i.e., as
defining a bundle over $\hat{M}$), or vice versa. But this might
very well happen, because in general
$H^1(M_{\text{red}},\Cal{O}_{\text{ev}}^{\times})$ and
$H^1(M_{\text{red}},\hat{\Cal{O}}_{\text{ev}}^{\times})$ are both
quotients of a free rank $g$ $\Lambda$-module (where $g$ is the genus of
$M_{\text{red}}$, see \cite{BR} for the calculation of these dimensions),
and the two quotients {\it are different in general\/}. Moreover, this also
shows that the bundles allowed in a nontrivial decomposition of
$\Cal{O}_2$ are necessarily flat (hence of degree zero).

\vskip 6pt
Finally, let us point out that  one can gain some further insight into the
meaning of this cohomological obstruction by considering the relatively
simple but important  case of  generic SKP curves (which
are the curves needed for studying the super KP equation); these are curves
for which the structure sheaf of the associated split curve has the form
$\Cal{O}_{\text{red}}|\Cal{N}$ (recall that a vertical bar means ``direct
sum of even and odd parts''), with
$\deg\Cal{N}=0$, but $\Cal{N}\neq\Cal{O}_{\text{red}}$. A simplifying
feature of these curves is that they have free cohomology, so in this case
$h^1(M,\Cal{O})=g\vert g-1$,  while, as already
mentioned,
$H^1(M_2,\Cal{O}_2)$ and $H^1(\hat{M},\hat\Cal{O})$ are in general only 
quotients of free $\Lambda$-modules, of ranks $g+1\vert
g-1$ and
$g|0$ respectively.

The space of line bundles of degree zero on $M$ can be identified with
the quotient $H^1(M, \Cal{O}_{\text{ev}}) / H^1(M_{\text{red}}, \Bbb{Z})$,
and similarly for $\hat{M}$ and $M_2$.
Note that elements of $H^1(M, \Cal{O}_{\text{ev}})$ are linear
combinations of both the $g$ even generators of
$H^1(M,\Cal{O})$ with coefficients from $\Lambda_{\text{ev}}$, and 
the $g-1$ odd generators with coefficients from $\Lambda_{\text{odd}}$.
On $M_{\text{red}}$ the space of degree-zero bundles is $g$-dimensional,
and they can all be given by constant $({\Bbb C}^\times$-valued)
transition functions.
Correspondingly, the bundles having constant 
($\Lambda_{\text{ev}}^\times$-valued) transition functions account for
the even generators of $H^1(M,\Cal{O})$ and all of
$H^1(\hat{M},\hat{\Cal{O}})$. 
Lifting bundles of degree zero from $M$ to
$M_2$ accounts for at most those bundles coming from (a quotient of) a
rank $g|g-1$
$\Lambda$-module in $H^1(M_2, \Cal{O}_2)$, and lifting the degree-zero
bundles from $\hat{M}$ adds nothing new.
Therefore there is an ``extra" bundle on $M_2$ which does not come from
lifting bundles of degree zero on $M$ or $\hat{M}$, nor does it
factor into these.
It can be viewed as the generator of the group $H^1(M,\Cal{B}er)$
in the long cohomology sequence associated to (1).
This group has rank $0|1$, being Serre dual to $H^0(M,\Cal{O})$ in
the category of supercurves.

To identify this
bundle, consider a covering of
$M_{\text{red}}$ consisting of an open disk $D$ centered at a point $P$,
with coordinate
$z=0$, and
$M_{\text{red}}\setminus\{P\}$; then the extra line bundle has as
transition function in the annulus
$$1-k\frac{\theta\rho}{z}=\frac{1}{z^k}(z-\theta\rho)^k.\tag{4}$$
 That this has to be the form of the transition function can be justified
by the fact that the residue mapping is integration of the principal parts
representing elements of
$H^1(M,\Cal{B}er)$, so the dual of a constant
function
$\kappa$ should represent a bundle having a transition function with a pole
of the form
$\kappa/z$ (this is an odd morphism, which explains why we get even
bundles; again see
\cite{BR} for the details on these constructions).

The point is that when
$k$ is not an integer,
$z^k$ {\it is not\/} single valued in
$D\setminus\{P\}$ so, in these cases, the functions appearing in the
decomposition given by the right hand side of (4) cannot define line
bundles over
$M$ and
$\hat{M}$, and  one sees that some (indeed,
most) of these bundles cannot come from tensor products of bundles. Also,
observe that the $N=2$ bundles appearing in this situation all have degree
0, but in case $k$ is an integer, this gives a decomposition into
$N=1$ bundles of degrees $k$ and $-k$.

\vskip 30pt
{\bf Acknowledgement.} F. Ongay wishes to warmly thank the Department of
Mathematics at UCSD, where the first version of this work was written, for
its kind hospitality.

\enddocument